\documentclass[11pt]{article}

\usepackage[utf8]{inputenc}
\usepackage[T1]{fontenc}
\usepackage{amsmath, amsfonts, amssymb}
\usepackage{graphicx}
\usepackage{hyperref}
\usepackage{geometry}
\usepackage{natbib}
\usepackage{booktabs}
\usepackage{float}

\hypersetup{
    colorlinks=true,
    linkcolor=blue,
    filecolor=magenta,      
    urlcolor=cyan,
    citecolor=red
}

\title{Fast Compute via MC Boosting}
\author{Sarah Polson\thanks{Oxford University. Email: \texttt{sarahmpolson@gmail.com}.}\\Oxford University
\and 
Vadim Sokolov\thanks{George Mason University. Associate Professor in Operations Research. Email: \texttt{vsokolov@gmu.edu}.}\\George Mason University}
\date{\today}

\begin{document}

\maketitle

\begin{abstract}
Modern training and inference pipelines in statistical learning and deep learning repeatedly invoke linear-system solves as inner loops, yet high-accuracy deterministic solvers can be prohibitively expensive when solves must be repeated many times or when only partial information (selected components or linear functionals) is required. We position \emph{Monte Carlo boosting} as a practical alternative in this regime, surveying random-walk estimators and sequential residual correction in a unified notation (Neumann-series representation, forward/adjoint estimators, and Halton-style sequential correction), with extensions to overdetermined/least-squares problems and connections to IRLS-style updates in data augmentation and EM/ECM algorithms. Empirically, we compare Jacobi and Gauss--Seidel iterations with plain Monte Carlo, exact sequential Monte Carlo, and a subsampled sequential variant, illustrating scaling regimes that motivate when Monte Carlo boosting can be an enabling compute primitive for modern statistical learning workflows.
\end{abstract}


\newpage

\section{Introduction}
\label{sec:introduction}

The need for fast computation has become increasingly important in modern statistical analysis, statistical learning, and deep learning applications. As datasets grow larger and models become more complex, traditional computational methods often become prohibitively slow.

In statistical learning and deep learning, a recurring bottleneck is the need to solve linear systems inside training and inference procedures. Examples include: (i) least squares and ridge-type updates in alternating minimization and block-coordinate methods; (ii) approximate Bayesian inference based on local Gaussian structure (e.g., Laplace approximations), which repeatedly requires solving systems involving Hessians or Fisher/Gauss--Newton surrogates; (iii) kernel methods and Gaussian-process inference, where prediction and hyperparameter updates require solving linear systems with kernel matrices; (iv) state-space models (Kalman filtering/smoothing) whose core updates are linear solves; and (v) implicit differentiation and equilibrium/implicit layers, where gradients require solving linear systems involving Jacobians.
In many of these settings, the dominant computational cost is not the model evaluation but the repeated linear solves, which can prevent otherwise attractive algorithms from being deployed at scale. Crucially, many pipelines require only \emph{selected components} of a solution or \emph{linear functionals} (e.g., predictions at a subset of test points, a preconditioned direction, or a small coordinate block updated in an outer loop). Deterministic factorization-based solvers typically compute the full solution, paying $O(m^3)$ (or $O(m^2)$ per iteration) even when only a small subset of outputs is required. Monte Carlo boosting targets this regime: it directly estimates components or functionals, is naturally parallelizable, and provides an explicit accuracy--compute trade-off through sampling and sequential residual correction.

A representative example is curvature-aware post-training quantization (e.g., GPTQ \citep{frantar2023gptq}, building on ``optimal brain'' analyses \citep{lecun1990optimal,hassibi1993obs}), where second-order information is used to compensate quantization error. In these approaches, quantization of a weight block $w$ is guided by a quadratic proxy
\begin{equation}
    \Delta \mathcal{L} \approx \frac{1}{2} (w - q(w))^\top H (w - q(w)),
\end{equation}
where $H$ is a (block) Hessian or second-moment matrix estimated from calibration activations. The resulting update rules require repeated access to inverse actions $(H+\lambda I)^{-1}v$ or to selected inverse columns for sequential error compensation. This is a natural setting for componentwise randomized estimators and sequential correction.
Recent work on inverse-Hessian approximations for compression (e.g., WoodFisher \citep{singh2020woodfisher}) further underscores the centrality of linear solves in modern compression pipelines.

This paper is intended as a computational-mathematics and statistics-facing synthesis of Monte Carlo approaches to linear systems. Our goal is not to introduce a new class of estimators, but rather to (i) place several related random-walk constructions in a common notation, (ii) clarify when each estimator is appropriate (full solution vs.\ selected components vs.\ linear functionals), and (iii) provide a small set of reproducible numerical illustrations that emphasize scaling behavior and practical trade-offs. We use the term \emph{Monte Carlo boosting} to refer to Halton's sequential Monte Carlo correction scheme, which repeatedly estimates and corrects residual systems to accelerate a base Monte Carlo estimator.

The main contributions are:
\begin{enumerate}
    \item A unified presentation of Neumann-series fixed-point formulations for linear systems arising from discretized differential and integral equations.
    \item A compact derivation of forward and adjoint random-walk estimators for linear functionals and solution components, together with a discussion of when each is computationally preferable.
    \item An explicit connection between Halton-style sequential correction (``boosting'') and weighted least squares / IRLS updates that arise in data augmentation and EM/ECM algorithms for non-Gaussian models.
    \item Empirical comparisons (with code) illustrating distinct scaling regimes across Jacobi/Gauss--Seidel, plain Monte Carlo, sequential Monte Carlo, and a subsampled sequential variant.
\end{enumerate}

\subsection{Connection to Existing Literature}

Monte Carlo methods originate in the foundational work of \citet{metropolis1949monte} and are treated in classic references such as \citet{hammersley1964monte} and historical surveys such as \citet{halton1970retrospective}. In numerical linear algebra, deterministic algorithms for linear systems and least squares are standard \citep{golub2013matrix}; the focus here is on regimes where randomized estimators can be competitive because they target \emph{components} or \emph{functionals} of the solution and can trade accuracy for compute.

The specific random-walk constructions reviewed in this paper can be traced to early stochastic representations of inverse operators and Neumann-series expansions. The iteration-of-operators perspective of \citet{curtiss1953monte} formalizes Monte Carlo estimation for linear operators, while \citet{forsythe1950matrix} develops matrix inversion by a Monte Carlo method using random-walk ideas. Closely related approaches were also studied for integral equations \citep{cutkosky1951monte}, which provide a natural motivation for random-walk estimators after discretization.

Sequential correction methods for linear systems were introduced by \citet{halton1962sequential} and subsequently developed for large-scale settings \citep{halton1994sequential}, with a practical summary in \citet{halton2008practical}. More broadly, Monte Carlo algorithms for matrix computations and their implementation considerations are discussed in \citet{dimov1998monte}. The present paper also emphasizes two themes that complement plain random-walk estimators in practice: variance reduction (e.g., residual Monte Carlo \citep{evans2003residual,evans2007residual}) and trajectory reuse across multiple unknowns or right-hand sides \citep{ji2012reusing}.

The remainder of the paper is organized as follows. Section~\ref{sec:setup} formulates the linear-system problem and its Neumann-series representation. Section~\ref{sec:random-walk} presents forward and adjoint random-walk estimators. Section~\ref{sec:boosting} discusses Monte Carlo boosting via sequential residual correction and connects the approach to IRLS-type updates that arise in data augmentation and EM/ECM algorithms. Section~\ref{sec:experiments} presents numerical illustrations and benchmarking. Section~\ref{sec:conclusion} concludes, and Appendix~\ref{sec:appendix} gives an integral-equation motivation.

\section{Problem Setup and Neumann Series}
\label{sec:setup}

Many problems in computational statistics and applied mathematics reduce to linear inverse problems, for example discretizations of Fredholm equations or elliptic boundary value problems. After discretization, these problems yield linear systems of the form $AX=B$ (possibly with multiple right-hand sides) or overdetermined least squares systems. The methods reviewed below are targeted at regimes where forming or factorizing $A$ is expensive, but where either (i) only a subset of components of $X$ are needed, or (ii) approximate solutions suffice inside an outer optimization/inference loop.

Algebraic equations often arise from discretizations of differential and integral equations. For differential equations, consider a Dirichlet operator:
\begin{equation}
    \left\{ \lambda(\xi, \eta) \frac{\partial^2}{\partial \xi^2} + \mu(\xi, \eta) \frac{\partial^2}{\partial \eta^2} \right\} X(\xi, \eta, \omega) = B(\xi, \eta, \omega).
\end{equation}
For integral equations of the second kind:
\begin{equation}
    \int_a^b A(\xi, \eta) X(\eta, \omega) \, d\eta = B(\xi, \omega),
\end{equation}
where $X(\xi, \eta, \omega)$ represents the continuous solution for the differential equation, and $X(\eta, \omega)$ for the integral equation. Discretization of these continuous problems leads to linear systems that must be solved numerically.

\subsection{Solving Linear Systems}

The goal is to solve the linear system of equations
\begin{equation}
    AX = B
\end{equation}
where $A$ is an $m \times m$ matrix, $B$ is $m \times n$, and $X$ is the $m \times n$ solution matrix.

Direct methods, such as Gaussian and Gauss-Jordan elimination, LU, and Cholesky decomposition, typically take $O(m^3) + O(m^2 n)$ operations. Iterative methods like Jacobi or Gauss-Seidel take $O(m^2 n s)$ for $s$ iterations, which can be computationally laborious for large $m$.

Sequential Monte Carlo techniques \citep{halton1994sequential} take time $O((m + cn)sN)$ with $N$ samples of length $s$. To determine $n_c$ components (a subset of $c$ rows of $X$), $cN$ replaces $m^2$ in the computational complexity. 

Expected errors for standard Monte Carlo are of order $O(N^{-1/2})$. However, sequential methods converge much more rapidly, with errors of order $\kappa^N$ for $0 < \kappa < 1$. This convergence is often not guaranteed to be random polynomial (RP), as $\kappa$ is usually the second eigenvalue of a Markov transition kernel.

To solve the system for $X$, we select a non-singular $m \times m$ matrix $G$ (a Markov transition kernel) such that $G^{-1}$ exists. Let $L = GB$ and $H = I - GA$. This yields the fixed-point equation:
\begin{equation}
    X = L + HX
\end{equation}
The von Neumann series expansion provides the solution:
\begin{equation}
    X = \sum_{r=0}^\infty H^r L = L + HL + H^2 L + \dots
\end{equation}
which converges if the spectral norm $\rho_H < 1$. This leads to the iterative process $X^{(g+1)} = L + HX^{(g)}$.

A key insight for faster sequential methods is that if $Y$ is an estimate of $X$ and we write $X = Y + Z$, then $Z$ satisfies:
\begin{equation}
    Z = D + HZ, \quad \text{where } D = L + HY - Y
\end{equation}
Thus, $Z$ satisfies the same form of equation as $X$, but with $L$ replaced by the error $D$ made when applying $H$ to $Y$. 

\subsection{Detailed Formulation}

Following \cite{halton1994sequential}, we consider a nonsingular system of equations
\begin{equation}
    \sum_{s=1}^{N} A_{rs} x_s = b_r, \quad r \in S = \{1, 2, \ldots, N\},
\end{equation}
where the $A_{rs}$ and $b_r$ are given real numbers, and the $x_s$ are the unknowns to be estimated, forming the solution vector $\mathbf{x}$. The nonsingularity implies that the $A_{rs}$ form a matrix $\mathbf{A}$ whose columns are linearly independent, so that the system $\mathbf{Ax} = \mathbf{b}$ has the unique solution
\begin{equation}
    \mathbf{x} = \mathbf{A}^{-1}\mathbf{b}.
\end{equation}

If the system is put into the form
\begin{equation}
    \mathbf{x} = \mathbf{a} + \mathbf{Hx},
\end{equation}
where the spectral radius of $\mathbf{H}$ is less than one, we have the Neumann series
\begin{equation}
    \mathbf{x} = (\mathbf{I} - \mathbf{H})^{-1}\mathbf{a} = \mathbf{a} + \mathbf{Ha} + \mathbf{H}^2\mathbf{a} + \cdots + \mathbf{H}^m\mathbf{a} + \cdots = \sum_{h=0}^{\infty} \mathbf{H}^h \mathbf{a},
\end{equation}
and the Neumann series is absolutely convergent. If we write the partial sums as
\begin{equation}
    \mathbf{X}_m = \sum_{h=0}^{m} \mathbf{H}^h \mathbf{a},
\end{equation}
we have $\mathbf{X}_m \to \mathbf{x}$ componentwise as $m \to \infty$, and we can set up \textit{iterative schemes} such as
\begin{equation}
    \mathbf{X}_0 = \mathbf{a}, \quad \mathbf{X}_m = \mathbf{a} + \mathbf{H}\mathbf{X}_{m-1}, \quad m > 0.
\end{equation}

\paragraph{Overdetermined Systems and Least Squares.}
In many practical instances, the ``known'' quantities $A_{rs}$ and $b_r$, or $H_{rs}$ and $a_r$, are themselves the result of lengthy calculations. An important case is when we have more equations than unknowns, with experimental or other errors making the system \textit{inconsistent}, as well as \textit{overdetermined}. We then have a system of approximate equations
\begin{equation}
    \sum_{s=1}^{N} L_{us} x_s \approx f_u, \quad u \in T = \{1, 2, \ldots, Q\},
\end{equation}
with $Q > N$, and we still assume that the columns of the matrix $\mathbf{L}$ are linearly independent. If we decide to minimize the ``sum of squares''
\begin{equation}
    S(\mathbf{x}) = \sum_{u=1}^{Q} \omega_u \left\{ f_u - \sum_{s=1}^{N} L_{us} x_s \right\}^2 = (\mathbf{f} - \mathbf{Lx})^\top \mathbf{\Omega} (\mathbf{f} - \mathbf{Lx}),
\end{equation}
where the $\omega_u > 0$ are diagonal elements of the diagonal matrix $\mathbf{\Omega}$, as our criterion of closeness of approximation, we obtain the weighted least squares solution
\begin{equation}
    \hat{\mathbf{x}} = (\mathbf{L}^\top \mathbf{\Omega} \mathbf{L})^{-1} \mathbf{L}^\top \mathbf{\Omega} \mathbf{f}.
\end{equation}
This formulation connects directly to iteratively reweighted least squares (IRLS) algorithms, where the weights $\omega_u$ are updated at each iteration based on the current residuals.

\section{Random-Walk Monte Carlo Estimators}
\label{sec:random-walk}

\subsection{Forward estimator for linear functionals}

The Monte Carlo approach constructs random walks to estimate the solution. Let the walk-related weight $W$ and the random variable $X$ be calculated as follows. Define
\begin{equation}
    W_\ell(\nu) = \frac{h_{k_0} H_{k_0, k_1} H_{k_1, k_2} \cdots H_{k_{\ell-1}, k_\ell}}{p_{k_0} P_{k_0, k_1} P_{k_1, k_2} \cdots P_{k_{\ell-1}, k_\ell}}, \quad \text{for } \ell = 0, 1, 2, \ldots,
\end{equation}
where $p_i$ is the initial probability and $P_{i,j}$ are transition probabilities. The estimator is
\begin{equation}
    X(\nu) = \sum_{\ell=0}^{\infty} W_\ell b_{k_\ell}.
\end{equation}
It can be shown \citep{dimov1998monte} that $\mathbb{E}[X] = \langle \mathbf{h}, \mathbf{x} \rangle$ and more specifically $\mathbb{E}[W_\ell f_{k_\ell}] = \langle \mathbf{h}, \mathbf{H}^\ell \mathbf{f} \rangle$. Since the random variable is an unbiased estimator of the solution, the MC method runs simulations of this random walk and uses the empirical mean $\bar{X}$ to approximate $\mathbb{E}[X]$.

If the desired output is the functional $\langle \mathbf{h}, \mathbf{x} \rangle$ itself, this method is effective. However, to estimate the entire solution vector, the method must be applied multiple times to obtain each entry of the solution. This limitation is addressed by the adjoint method.

\subsection{Adjoint method (many components from shared walks)}

\paragraph{Adjoint Method.}
In the forward method, the right-hand side $\mathbf{b}$ only arises in the final path-dependent random variable $X(\nu)$. We could easily change $\mathbf{b}$ to estimate solutions of slightly different linear systems. However, to vary the vector $\mathbf{h}$ (which determines which components of the solution we estimate), we would need to control the start of the walks. The adjoint method inverts these roles: it creates a method that \textit{starts} depending on $\mathbf{b}$ and estimates solution entries with path-dependent weights.

Consider the adjoint linear system
\begin{equation}
    \mathbf{y} = \mathbf{H}^\top \mathbf{y} + \mathbf{d}.
\end{equation}
If $\mathbf{x}$ solves $\mathbf{x} = \mathbf{a} + \mathbf{Hx}$, then we have the following inner product equivalence:
\begin{equation}
    \langle \mathbf{x}, \mathbf{d} \rangle = \langle (\mathbf{I} - \mathbf{H})^\top \mathbf{y}, \mathbf{x} \rangle = \langle \mathbf{y}, (\mathbf{I} - \mathbf{H})\mathbf{x} \rangle \quad \Longrightarrow \quad \langle \mathbf{x}, \mathbf{d} \rangle = \langle \mathbf{y}, \mathbf{b} \rangle.
\end{equation}
We can apply the forward method technique to compute the functional $\langle \mathbf{y}, \mathbf{b} \rangle$. This requires that walks start depending on $\mathbf{b}$. Then we can vary $\mathbf{d}$ through $\mathbf{e}_1, \ldots, \mathbf{e}_n$ depending on which state the random walk lands at each step. In this way, by a single random walk, each solution element $x_1, \ldots, x_n$ can be updated by different random walk steps.

The adjoint method constructs the following Markov chain with initial probability
\begin{equation}
    \Pr(X_0 = i) = p_i \quad \text{such that} \quad b_i \neq 0 \Rightarrow p_i \neq 0,
\end{equation}
and transition probability
\begin{equation}
    \Pr(X_{\ell+1} = j \mid X_\ell = i) = P_{i,j} \quad \text{such that} \quad H_{i,j}^\top \neq 0 \Rightarrow P_{i,j} \neq 0.
\end{equation}
Given a realization $\nu$, we define the weight for each step as
\begin{equation}
    W_\ell = \frac{h_{k_0} H_{k_0, k_1}^\top H_{k_1, k_2}^\top \cdots H_{k_{\ell-1}, k_\ell}^\top}{p_{k_0} P_{k_0, k_1} P_{k_1, k_2} \cdots P_{k_{\ell-1}, k_\ell}}, \quad \text{for } \ell = 0, 1, 2, \ldots
\end{equation}
This construction allows estimation of all solution components from a single set of random walks, providing substantial computational savings when the full solution vector is required.

\section{Monte Carlo Boosting and IRLS Connections}
\label{sec:boosting}

\subsection{Sequential residual correction (``boosting'')}

The Neumann-series fixed-point form $X = L + HX$ admits a residual-correction perspective that underlies Halton's sequential Monte Carlo method. If $Y$ is a current estimate of $X$ and we write $X = Y + Z$, then the correction $Z$ satisfies $Z = D + HZ$ with $D = L + HY - Y$. Monte Carlo boosting iteratively estimates $Z$ (or selected components of $Z$) and updates $Y \leftarrow Y + \widehat{Z}$, reusing the same operator $H$ but with a progressively smaller right-hand side $D$. In practice, this can dramatically reduce the number of Monte Carlo steps needed to reach a target accuracy \citep{halton1994sequential,halton2008practical}.

\subsection{Halton method with iteratively reweighted least squares}

Combining the Halton method with iteratively reweighted least squares (IRWLS) \citep{green1984iteratively} enables application to non-Gaussian likelihoods via data augmentation \citep{polson2012data}. In the EM framework, the M-step requires solving a linear system for the parameter update:
\begin{equation}
    \hat{\beta} = (\tau^{-2}\hat{\Lambda} + \mathbf{X}^\top \hat{\Omega} \mathbf{X})^{-1} (\mathbf{y}^* + \mathbf{b}^*),
\end{equation}
where $\hat{\Lambda}$ and $\hat{\Omega}$ are diagonal matrices of current shrinkage and precision estimates. This system can be solved efficiently using the Halton method described above.

\cite{polson2011data} applied this approach to Support Vector Machines (SVM), where the M-step is essentially a weighted least squares problem with weights $\lambda_i^{-1}$. A key computational advantage arises as coefficients approach zero: the corresponding rows and columns can be removed from the matrix $\mathbf{X}^\top \Lambda^{-1} \mathbf{X}$, progressively reducing the problem dimension.

\paragraph{Stable Version of the M-Step.}
Let $\mathbf{X}_{-s}$ denote $\mathbf{X}$ with the support vector rows deleted, let $\lambda_{-s}^{-1}$ denote the finite elements of $\lambda^{-1}$, and let $\Lambda_{-s}^{-1} = \text{diag}(\lambda_{-s}^{-1})$. A numerically stable version of the M-step can be formulated via restricted least squares using Lagrange multipliers and block matrix inverses; see, e.g., \citet{golub2013matrix}:
\begin{equation}
    \begin{pmatrix}
    \mathbf{X}_{-s}^\top (\mathbf{1} + \lambda_{-s}^{-1}) \\
    \mathbf{1}
    \end{pmatrix}
    =
    \begin{pmatrix}
    B_{-s} & \mathbf{X}_s^\top \\
    \mathbf{X}_s & 0
    \end{pmatrix}
    \begin{pmatrix}
    \beta \\
    \psi
    \end{pmatrix},
\end{equation}
where $\psi$ is a vector of Lagrange multipliers and
\begin{equation}
    B_{-s} = \nu^{-2} \Sigma^{-1} \Omega^{-1} + \mathbf{X}_{-s}^\top \Lambda_{-s}^{-1} \mathbf{X}_{-s}.
\end{equation}
The inverse of the partitioned matrix can be expressed in closed form:
\begin{equation}
    \begin{pmatrix}
    B_{-s}^{-1}(I + \mathbf{X}_s^\top F \mathbf{X}_s B_{-s}^{-1}) & -B_{-s}^{-1} \mathbf{X}_s^\top F \\
    -F \mathbf{X}_s B_{-s}^{-1} & F
    \end{pmatrix},
\end{equation}
where $F = -(\mathbf{X}_s B_{-s}^{-1} \mathbf{X}_s^\top)^{-1}$. This partitioned inverse provides a closed-form solution that avoids numerical instabilities when some weights become very large.

\section{Numerical Experiments}
\label{sec:experiments}

\subsection{Comparison of methods (complexity)}

Table~\ref{tab:methods} summarizes the computational complexity of different approaches for solving linear systems. For context, \emph{direct} methods (LU/Cholesky) factorize $A$ once and then solve by triangular substitution, delivering a high-accuracy solution but at cubic cost in $m$. \emph{Stationary iterations} (Jacobi, Gauss--Seidel) update an iterate by repeated matrix--vector operations; their cost scales with the number of iterations $s$ needed for convergence and depends strongly on conditioning. \emph{Plain Monte Carlo} uses random-walk estimators to target selected components or linear functionals, so the cost depends on the number of queried components $c$ and random walks $N$ rather than on forming full matrix factorizations. \emph{Sequential Monte Carlo} (Halton boosting) refines an initial Monte Carlo estimate by solving a sequence of residual systems; the ``exact'' variant uses full residual updates, while the \emph{sampled} variant reduces per-iteration cost by subsampling columns/coordinates during correction.

\begin{table}[H]
\centering
\caption{Comparison of methods for solving $AX = B$ where $A$ is $m \times m$.}
\label{tab:methods}
\begin{tabular}{@{}lll@{}}
\toprule
\textbf{Method} & \textbf{Complexity} & \textbf{Notes} \\
\midrule
Direct (LU, Cholesky) & $O(m^3)$ & Exact solution \\
Jacobi iteration & $O(m^2 s)$ & $s$ iterations, requires convergence \\
Gauss-Seidel & $O(m^2 s)$ & Faster convergence than Jacobi \\
Plain Monte Carlo & $O(cN)$ & $c$ components, $N$ random walks \\
Exact Sequential MC & $O(m^2 s)$ & Geometric convergence $\kappa^N$ \\
Sampled Sequential MC & $O(m s)$ & Subsampling reduces per-iteration cost \\
\bottomrule
\end{tabular}
\end{table}

The scaling study in the next subsection provides an empirical illustration of these regimes. In our benchmark, Gauss--Seidel exhibits the steepest growth as $m$ increases, reflecting the $O(m^2)$ cost of each iteration, while Jacobi shows similar quadratic behavior but with a smaller constant factor due to simpler updates. The exact sequential correction method achieves favorable constants by exploiting the geometric convergence described in Section~\ref{sec:setup}, and the sampled sequential variant reduces per-iteration cost by subsampling, leading to approximately linear scaling in $m$. Plain Monte Carlo is essentially flat across problem sizes when estimating a fixed number of components $c$, since its runtime depends primarily on the random-walk budget rather than the ambient dimension.

\subsection{Scaling study}

To demonstrate the computational advantages of Monte Carlo methods, we conduct a benchmark study comparing five solution methods across a range of problem sizes. We generate random diagonally dominant matrices of increasing dimension $m \in \{1000, 2000, 3000, 4000, 5000\}$. For each matrix size, we construct a system $AX = B$ where the matrix $A$ is designed to ensure convergence of iterative methods by maintaining a spectral radius $\rho_H < 1$.

\begin{figure}[H]
    \centering
    \includegraphics[width=0.7\textwidth]{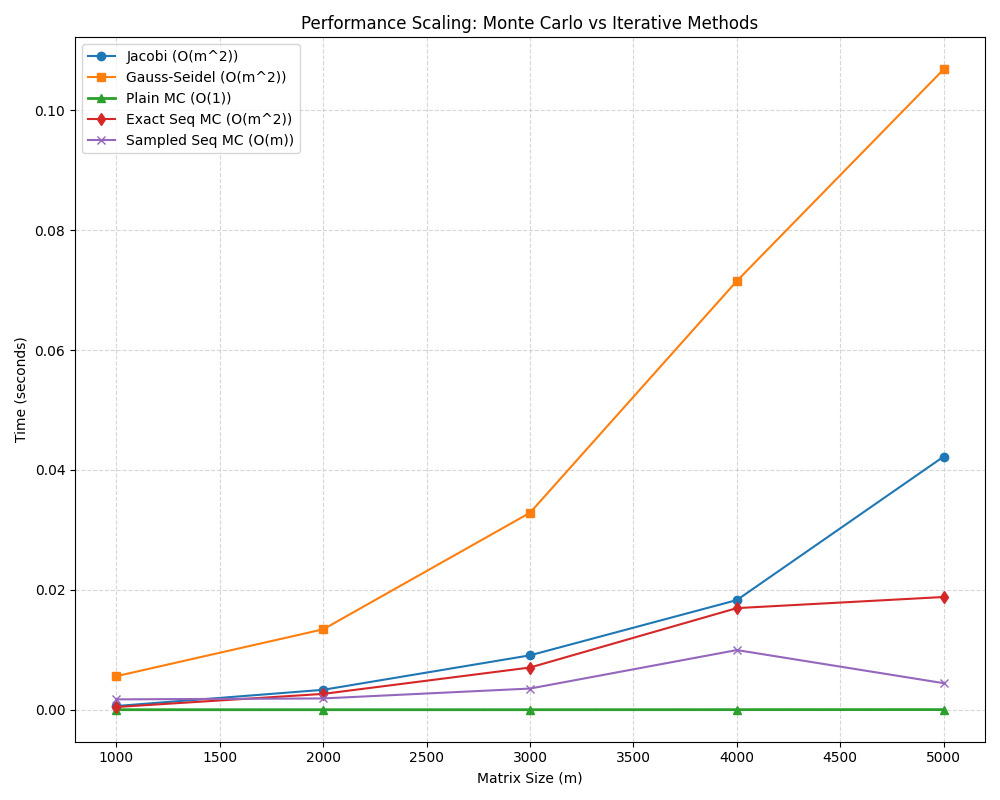}
    \caption{Runtime comparison of five methods for solving linear systems as matrix size $m$ increases from 1000 to 5000. Traditional iterative methods (Jacobi, Gauss-Seidel) exhibit $O(m^2)$ scaling. Sequential Monte Carlo methods achieve improved scaling through geometric convergence and subsampling. Plain Monte Carlo achieves $O(1)$ scaling in $m$ when estimating a fixed number of solution components.}
    \label{fig:performance}
\end{figure}

Figure~\ref{fig:performance} presents the runtime scaling results. The performance differences across methods are striking and confirm the theoretical complexity bounds derived above.

\paragraph{Reporting caveat.}
The timing results are intended to illustrate asymptotic scaling rather than to serve as a definitive performance contest: constants depend on implementation language, linear algebra backends, and stopping criteria. A submission-grade benchmarking study would additionally report accuracy (e.g., residual norms) at comparable solution quality across methods and provide full experimental details (hardware, BLAS/LAPACK versions, random seeds, and stopping rules). The present study is therefore best interpreted as an empirical complement to the complexity discussion in Table~\ref{tab:methods}.

Table~\ref{tab:timing} presents the raw timing data. The Plain Monte Carlo method shows times below measurable precision (reported as 0.0000), confirming its independence from matrix size. Gauss-Seidel times grow from 0.0056s to 0.1068s---a 19-fold increase---while Sampled Sequential MC grows only from 0.0017s to 0.0044s, a 2.6-fold increase consistent with $O(m)$ scaling.
At the largest problem size ($m=5000$), Sampled Sequential MC is approximately $0.1068/0.0044 \approx 24\times$ faster than Gauss--Seidel in this implementation, illustrating the regime where sequential correction with subsampling can be a practical win.

\begin{table}[H]
\centering
\caption{Runtime (seconds) for solving $AX = B$ across matrix sizes.}
\label{tab:timing}
\begin{tabular}{@{}rrrrrr@{}}
\toprule
$m$ & Jacobi & Gauss-Seidel & Plain MC & Exact Seq MC & Sampled Seq MC \\
\midrule
1000 & 0.0006 & 0.0056 & 0.0000 & 0.0005 & 0.0017 \\
2000 & 0.0033 & 0.0134 & 0.0000 & 0.0026 & 0.0019 \\
3000 & 0.0091 & 0.0329 & 0.0000 & 0.0070 & 0.0035 \\
4000 & 0.0183 & 0.0715 & 0.0000 & 0.0170 & 0.0100 \\
5000 & 0.0422 & 0.1068 & 0.0000 & 0.0188 & 0.0044 \\
\bottomrule
\end{tabular}
\end{table}

\subsection{Monte Carlo convergence illustration (variance caution)}

To illustrate Monte Carlo convergence directly, we run the R implementation of Halton's direct estimator in \texttt{code/mc-linear-sys.R} on a fixed problem of size $m=1000$. We generate a dense matrix $H$ with entries $H_{ij} = 0.9/(m+i+j)$ and a known solution $x$; we then form $a = x - Hx$ so that $(I-H)x = a$. For each Monte Carlo budget $M$, we generate $M$ independent random-walk estimates and compute the averaged estimator $\hat{x}_M$.

\begin{figure}[H]
    \centering
    \includegraphics[width=0.6\textwidth]{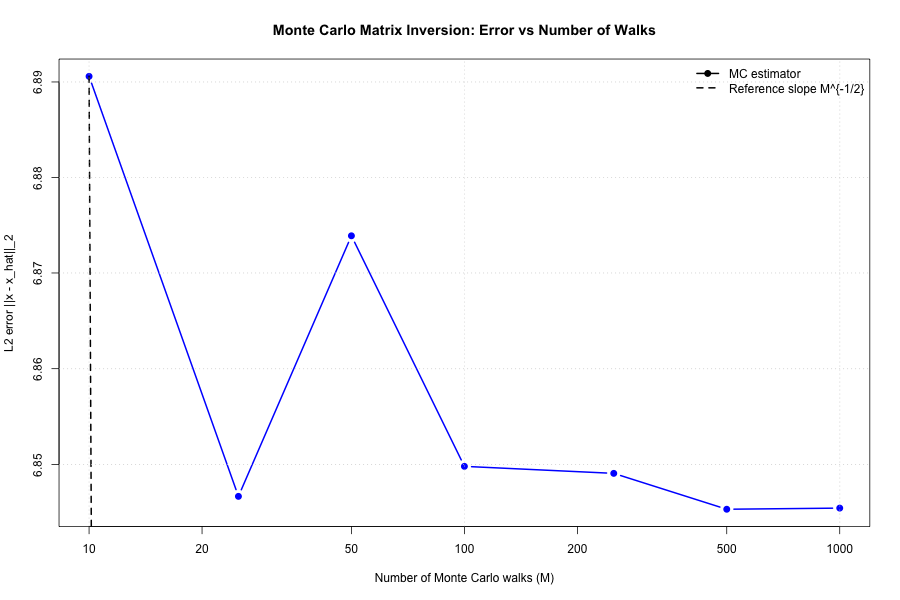}
    \caption{Monte Carlo convergence for Halton's direct estimator on a fixed $m=1000$ problem. The dashed line indicates a reference $M^{-1/2}$ rate.}
    \label{fig:mcconv}
\end{figure}

Table~\ref{tab:mcconv} reports the empirical $\ell_2$ error $\|x-\hat{x}_M\|_2$ and the runtime required to generate $M$ random-walk draws. Figure~\ref{fig:mcconv} plots the error versus $M$ on a log-log scale, together with a reference $M^{-1/2}$ slope. In this parameterization, the estimator exhibits substantial variance and the error decreases slowly with $M$.
We emphasize that this experiment is a \emph{cautionary illustration} rather than a performance claim: for dense, poorly preconditioned systems and naive transition kernels, plain random-walk estimators can have very high variance.
This motivates sequential correction (Halton boosting), adjoint estimators that reuse trajectories across components, and variance reduction techniques such as residual Monte Carlo \citep{evans2003residual,evans2007residual} and walk reuse \citep{ji2012reusing}.

\begin{table}[H]
\centering
\caption{Monte Carlo convergence for the linear system experiment in \texttt{code/mc-linear-sys.R} ($m=1000$).}
\label{tab:mcconv}
\begin{tabular}{@{}rrr@{}}
\toprule
$M$ & $\|x-\hat{x}_M\|_2$ & Time (s) \\
\midrule
10   & 6.8906 & 0.002 \\
25   & 6.8467 & 0.003 \\
50   & 6.8739 & 0.005 \\
100  & 6.8498 & 0.009 \\
250  & 6.8491 & 0.025 \\
500  & 6.8453 & 0.050 \\
1000 & 6.8455 & 0.102 \\
\bottomrule
\end{tabular}
\end{table}

\section{Discussion}
\label{sec:discussion}

The numerical results presented above illustrate the central thesis of this paper: Monte Carlo methods can dramatically reduce computational cost for solving linear systems when the problem structure permits. The key insight from Halton's sequential method is that the error correction equation $Z = D + HZ$ has the same form as the original problem, enabling iterative refinement with geometrically decreasing errors.

Several practical considerations guide the choice of method. First, when only a subset of solution components is required, Plain Monte Carlo offers unparalleled efficiency. Second, when the spectral radius $\rho_H$ is close to unity, the geometric convergence rate $\kappa^N$ may be slow, favoring traditional methods. Third, the Sampled Sequential MC approach provides a middle ground, trading some accuracy for linear rather than quadratic scaling.

Variance reduction ideas such as residual Monte Carlo provide complementary strategies for improving estimator efficiency \citep{evans2003residual,evans2007residual}. Recent work also shows that substantial speedups can be obtained by reusing random walks across multiple unknowns in Monte Carlo linear solvers \citep{ji2012reusing}. Connections to reinforcement learning and stochastic control arise naturally when randomized operators are used for value-function approximation and policy improvement \citep{barto1993monte,polson2011simulationbased,pmlr-vR3-duff01a,whittle1988policy,whittle1988restless}. For sequential latent-variable estimation, online EM provides an alternative incremental perspective \citep{cappe2009online}.

Future work will extend these methods to structured matrices arising in statistical learning and deep learning applications, including kernel matrices and attention mechanisms in transformer architectures, and to nonlinear systems where sequential Monte Carlo variants can be applied \citep{halton2006sequential}.

\section{Conclusion}
\label{sec:conclusion}
We reviewed Monte Carlo estimators for linear systems based on Neumann-series representations and random-walk constructions, with an emphasis on sequential correction (``boosting'') and its relevance for repeated linear solves inside statistical inference algorithms. The numerical illustrations highlight the regimes where Monte Carlo methods can be attractive (componentwise estimation, large-scale problems, and settings where approximate solves suffice), as well as the central limitation of naive estimators (variance) that motivates sequential correction and variance reduction. An important direction for future work is benchmarking at matched accuracy on application-driven problems (e.g., IRLS and data augmentation) where the cost of repeated deterministic solves dominates.

\appendix
\section{Appendix: Fredholm Integral Equations and Random-Walk Estimators}
\label{sec:appendix}

Monte Carlo methods for linear inverse problems can be motivated through integral equations. Consider a Fredholm equation of the first kind,
\begin{equation}
    f(y) = \int h(x,y)\, g(x)\, dx,
\end{equation}
which is typically discretized for computation into a linear system $f = Hg$. Such inverse problems are often ill-conditioned: small perturbations in the observed data $f$ can produce large changes in the recovered $g$. A statistical formulation addresses this by specifying a likelihood and prior, for example
\begin{equation}
    f \mid g \sim \mathcal{N}(Hg, \sigma^2 I), \qquad g \sim \pi(g),
\end{equation}
so that inference is based on the posterior $\pi(g\mid f)$ and point estimates can be obtained via posterior modes. In many settings, Monte Carlo algorithms for exploring $\pi(g\mid f)$ are natural competitors to EM-type procedures, and the choice of prior $\pi(g)$ plays an important regularizing role (e.g., Markov random fields or conditionally autoregressive priors).

This viewpoint is closely related to statistical treatments of ill-conditioned inverse problems and their computational implications; see, for example, \citet{polson1993discussion}.

Monte Carlo constructions are particularly transparent for Fredholm equations of the second kind,
\begin{equation}
    f(x) = g(x) + \int K(x,y)\, f(y)\, dy,
\end{equation}
which admit a Neumann-series solution when the integral operator has norm strictly less than one:
\begin{equation}
    f = g + K g + K^2 g + \cdots.
\end{equation}
This representation suggests a stochastic solution via a random walk. Let $P(\cdot,\cdot)$ be a strictly sub-stochastic transition kernel satisfying $\int P(x,y)\,dy < 1-\delta$ for some $\delta>0$, and assume $P$ has support contained in that of $K$. Consider a Markov chain $(X_0,X_1,\ldots)$ with transition kernel $P$, stopped at a random time $\tau$ (a ``killing'' time) defined by terminating at step $n$ with probability
\begin{equation}
    p(X_n) = 1 - \int P(X_n,y)\,dy.
\end{equation}
Define the incremental weight
\begin{equation}
    V(x,y) = \frac{K(x,y)}{P(x,y)},
\end{equation}
and the path-dependent estimator
\begin{equation}
    \widehat{f}(X_0) \;=\; \frac{g(X_\tau)}{p(X_\tau)} \prod_{i=1}^{\tau} V(X_{i-1},X_i).
\end{equation}
Under standard regularity conditions ensuring integrability of the weights, $\widehat{f}(X_0)$ is an unbiased estimator of $f(X_0)$. Moreover, explicit expressions for the variance can be derived, and many assumptions (including those involving the norm of $P$) can be weakened. The same construction extends to local estimators of inverse problems and suggests parallel implementations by running many independent random walks.

\bibliographystyle{apalike}
\bibliography{FastCompute}

\end{document}